\documentclass[twocolumn]{aastex62}

\graphicspath{{./}{figures/}}

\begin{document}


\title[Improved parameters for HATS-24b]{First observation of a planetary transit with the SPARC4 CCD: improved parameters for HATS-24b}

\correspondingauthor{Janderson M. Oliveira}
\email{jandersonfba.unifei@gmail.com, joliveira@lna.br}

\author[0000-0002-5171-4842]{Janderson M. Oliveira}
\affiliation{Universidade Federal de Itajub\'{a} (UNIFEI), Av. BPS 1303, Itajub\'{a}, MG 37500-903, Brazil}

\author[0000-0002-5084-168X]{Eder Martioli}
\affiliation{Laborat\'{o}rio Nacional de Astrof\'{i}sica (LNA/MCTIC), Rua Estados Unidos 154, Itajub\'{a}, MG 37504-364, Brazil}

\author[0000-0002-0016-8597]{Marcelo Tucci-Maia}
\affiliation{Universidad Diego Portales (UDP), Manuel Rodr\'{i}guez Sur 415, Regi\'{o}n Metropolitana, Santiago, Chile}



\begin{abstract}

The photometric monitoring of planetary transits is an important method for the characterization of exoplanets. Transiting exoplanets allow measurements of the planetary radius and mass, providing information on the physical structure of the planet. Therefore, the parameters obtained with the transit method become fundamental for a comparative study of exoplanets in different planetary systems. In this work, we present the results obtained from observations of a transit event of the exoplanet HATS-24b using the new SPARC4 CCD mounted on the 1.6 m telescope of the Pico dos Dias Observatory (OPD). We have used Bayesian statistical inference to determine the physical parameters of this exoplanet, where we obtained a radius of 1.395$\pm$0.057 $R_{\rm J}$, mass of 2.26$\pm$0.17 $M_{\rm J}$, and density of 1.03$\pm$0.15 g\,cm$^{-3}$. We have combined our measurements of the central time of transit and orbital period with values from the literature to obtain an improved ephemeris for the transits of HATS-24b, $T_{c} = (2457948.709321 \pm 0.000039) + E (1.3484978 \pm 0.0000009)$. We have applied the differential analysis using a solar spectrum to recalculate the stellar surface parameters of HATS-24, where we obtained $T_{\rm eff}$ = $6125\pm94$ K, $\log{g}=4.370\pm0.045$~cm~s$ˆ{-2}$, and [Fe/H]=$-0.229\pm0.058$~dex. This allowed us to estimate an equilibrium temperature for HATS-24b of $T_{\rm eq}=2166\pm53$ K. The mass and radius of HATS-24b are consistent with the theoretical model of a pure helium-hydrogen planet at 1~Gy.

\end{abstract}



\section{Introduction} \label{sec:intro}

Currently, there are about 3800 exoplanets detected according to the NASA Exoplanet Archive catalog  \footnote{\url{https://exoplanetarchive.ipac.caltech.edu/}} \citep{Akeson2013} of which about 3000 of these objects are transiting exoplanets.  The majority of these transiting exoplanets have been detected by the Kepler mission \citep{Borucki2003}, which aimed at detecting terrestrial-type planets orbiting in the habitable zone of their host star. With the current ground-based instrumentation it is only possible to detect the transits of large planets in close-in orbits, which amounts a small fraction of all known transiting exoplanets. However, some of these systems still require follow-up observations to improve the characterization of the physical parameters of the exoplanets. The observations of several transit events can also improve the determination of the orbital period and allows one to search for transit time variations, which may lead to the detection of additional planets in the system.

Therefore, the class of planets we are particularly interested are the hot Jupiters, which are planets likely formed in the outer regions of their protoplanetary discs and that migrated later to the inner regions, reaching orbits within approximately 0.1~AU of their host star \citep{FoggNelson2006}. The hot Jupiters have orbital periods of up to 10 days \citep{Wang2015}, which implies the transit duration of a few hours, allowing the observation of a full transit event during the night. The probability of occurrence of transits becomes smaller for longer period planets due to geometric characteristics of the system \citep{Heller2009}.

The hot Jupiter HATS-24b was recently discovered by \cite{Bento2017}, through the Hungarian-made Automated Telescope Network-South survey (HATSouth) \footnote{\url{https://hatsouth.org/}} \citep{Bakos2013}. According to \cite{Bento2017}, HATS-24b orbits a G or F type star, with magnitude V = 12.830 and effective temperature of 5800$\pm$300\,K. \cite{Bento2017} measured a mass of HATS-24~b of 2.39 $_{-0.12}^{+0.21}\,M_{\rm J}$ and a radius of 1.516$_{-0.065}^{+0.085}\,R_{\rm J}$, resulting in a density of 0.92$\pm$0.15~g~cm$^{-3}$. The small amount of observations available in the literature and the low photometric precision of transits observed for this object have led us to select HATS-24b as a potential candidate for further follow-up. We were also motivated by the arrival of the new CCDs of the instrument Simultaneous Polarimeter and Rapid Camera in 4 bands (SPARC4) \citep{rodrigues12}. These CCDs have already been extensively tested in the lab by \cite{bernardes2018} and needed to be tested on the telescope to verify the real photometric capabilities of this new instrument. SPARC4 is under construction and will be installed on the 1.6~m Perkin-Elmer telescope at the Pico dos Dias Observatory (OPD), Brazil. This is expected to be the workhorse instrument of OPD, and will be an excellent instrument to follow-up transits of hot Jupiters, where it is particularly interesting because it will operate in four photometric bands simultaneously, allowing one to constrain the transit depth at different wavelengths. 

\section{Observations}
\label{sec:observations}

We have used the Exoplanet Transit Database - ETD\footnote{\url{http://var2.astro.cz/ETD/}} \citep{poddany2010} to predict the transit events of HATS-24b, which would be observable at OPD during the first semester of 2017.  We have been granted time (project number OP2017A-026) to use the 1.6~m telescope to observe the transit of HATS-24\,b on the night of July 13, 2017. We requested the new iXon Ultra 888 CCD of the SPARC4 project and the focal reducer. This set-up provides a field-of-view of 6.15' x 6.15', which allows simultaneous observation of several comparison stars in the field (see Fig.~\ref{fig:aperture}). We performed continuous time series photometry in the I-filter from the start of the night until the target reached airmass above the observable limit. We started obsevations at about 2 hours and 46 minutes before the predicted start of the transit and ended at about 1 hour and 36 minutes after the predicted end of the transit. Initially, we used 20 seconds of exposure time, but we changed to 30 seconds after 30 minutes of obsevations to improve signal-to-noise. The observations were performed under near photometric conditions.

\section{Data Reduction}
\label{sec:datareduc}

We have used the software AstroImageJ\footnote{\url{http://www.astro.louisville.edu/software/astroimagej/}} (AIJ) \citep{collins2017} to reduce our data. We have obtained 30 bias exposures and 20 flat-field exposures.  The AIJ software performs the bias subtraction and division by flat-field on each one of the 806 science exposures, where the bias and flat-field exposures were combined by the median.  The AIJ performs flux extraction and calculates the differential photometry for multiple apertures, providing differential light curves for each comparison star with respect to the target. Although AIJ allows the analysis of the transit light curve, we performed the analysis using a software tool developed by us. Our tool can fit a model that accounts for the transit model and for trends in the light curves of each individual comparison star. In addition, our analysis implements Bayesian statistical inference to estimate the model parameters.

\begin{figure}
\centering
\includegraphics[width=\columnwidth]{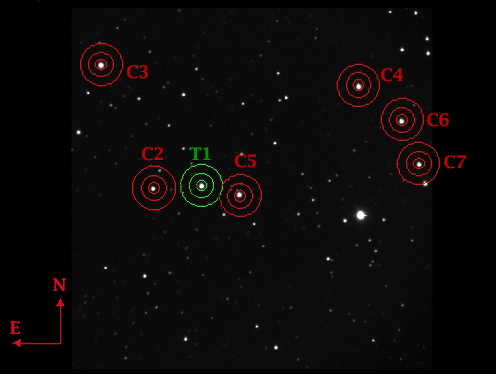}
\caption{Sample image of the field-of-view observed (6.15' x 6.15'). HATS-24 is labeled as T1 (in green), and the 6 comparison stars are labeled as C$n$ where $n = 2, .., 7$ (in red). For each star we present the circular apertures used for the star flux extraction and the concentric apertures used for the sky flux extraction.}
\label{fig:aperture}
\end{figure}

Fig.~\ref{fig:aperture}\ presents a sample image of the field-of-view observed, where the green circles show the apertures used to extract the flux of our target (HATS-24, labeled as T1) and red circles show the apertures used for the comparison stars. We use an aperture radius of 11 pixels to extract the flux and an aperture between 25 pixels (inner radius) and 50 pixels (outer radius) to extract the sky flux. We have selected the six comparison stars based on the similarity of their magnitudes to the target ($\pm$1.0 mag) as presented in Table~\ref{table:targets}. We have also used the Multi-Aperture (MA) mode of the AIJ, which recalculates the centroid of the target and comparison stars for each individual image of the time series to allow for centering corrections.


\begin{table*}
\centering
\caption{Target and comparison stars used in the differential photometry.}
\label{table:targets}
\begin{tabular}{ccccc}
\hline
Index & ID          & RA            & DEC            & I-MAG        \\
& (NOMAD-1)   & (hh:mm:ss.sss)  & ($\pm$dd:mm:ss.ss)  & (USNO-B1.0)  \\
\hline
T1  & 0282-0888453  & 17:55:33.779  &  -61:44:50.38  & 12.03        \\
C2  & 0282-0888407  & 17:55:28.688  &  -61:44:59.60  & 11.90        \\
C3  & 0282-0888506  & 17:55:40.194    &  -61:44:52.02  & 12.86        \\
C4  & 0282-0888552  & 17:55:46.887  &  -61:42:54.24  & 11.66        \\
C5  & 0282-0888299  & 17:55:12.608  &  -61:43:18.20  & 11.70        \\
C6  & 0282-0888251  & 17:55:06.769  &  -61:43:51.54  & 12.26        \\
C7  & 0282-0888235  & 17:55:04.544  &  -61:44:32.88  & 12.35        \\
\hline
\end{tabular}
\end{table*}

\section{Analysis}
\label{sec:analysis}

We performed our analysis using an adapted version of the software developed by \cite{martioli2018}, which was initially developed for the study of secondary eclipses. We made the necessary modifications in the code to apply the same methodology for the study of primary transits. This software, written in \texttt{Python}, uses the \texttt{BATMAN} package developed by \cite{Kreidberg2015}, which implements the geometric model of \cite{mandel&agol2002} for the calculation of the light curve for the transit model. It uses Bayesian inference to estimate the posterior probability of the transit model parameters and also the trend model coefficients for each comparison star. It uses the package EMCEE \citep{Foreman-Mackey2013} to generate samples based on a priori information of the physical parameters together with the data obtained. The \texttt{EMCEE} package uses the Markov Chain Monte Carlo sampler (MCMC) applying the method proposed by \cite{Goodman&Weare2010}. We use the parameters obtained by \cite{Bento2017} as a priori information as presented in Table~\ref{table:priors}. A normal probability distribution was adopted for the following parameters: central time of transit ($T_{c}$), orbital period ($P$), planet-to-star radius ($R_{p}/R_{\star}$),  semi-major axis ($a_{p}/R_{\star}$), orbital inclination ($i$), and limb darkening coefficients ($u_{1}$ and $u_{2}$), where we assume a quadratic limb darkening \citep{Kreidberg2015}. The following parameters are fixed as constant values: eccentricity ($e=0$) and longitude of periastron ($\omega=90^{\circ}$).

\begin{table}
\caption{Priors obtained from \citet{Bento2017}.}
\label{table:priors}
\begin{tabular}{cccc}
\hline
Parameter & Prior Type & Initial guess & Unit \\
\hline
$T_{c}$     & Normal     & 2457948.708      $\pm$ 0.010    & BJD \\
$P$      & Normal     & 1.3484954 $\pm$ 1.3e-06 & days \\
$R_{p}/R_{\star}$ & Normal     & 0.1307    $\pm$ 0.003   & - \\
$a_{p}/R_{\star}$  & Normal     & 4.67      $\pm$ 0.1     & - \\
$i$     & Normal     & 86.6      $\pm$ 1.2     & degrees \\
$e$     & Fixed      & 0.                      & degrees \\
$\omega$  & Fixed      & 90.                     & degrees \\
$u_{1}$           & Normal     & 0.1919    $\pm$ 0.0100  & - \\
$u_{2}$           & Normal     & 0.3654    $\pm$ 0.0100  & - \\
\hline
\end{tabular}
\end{table}

\section{Results}
\label{sec:results}

\subsection{Transit Parameters}
\label{sec:transitpars}

\begin{figure}
\includegraphics[width=0.45\textwidth]{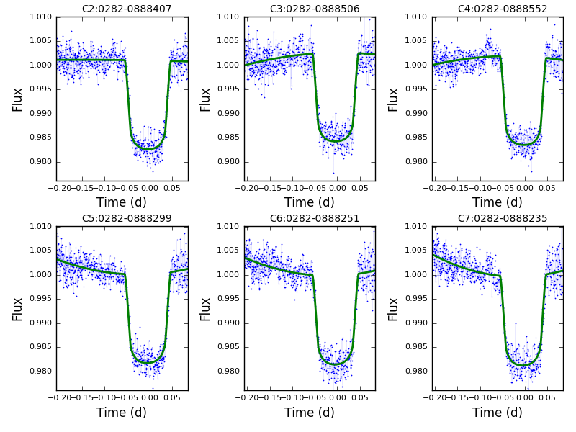}
\caption{Each panel shows the differential light curve (blue dots) for each comparison star and the fit model (green line), where the model includes the trend and the global transit model.}
\label{fig:starcomp}
\end{figure}

For each comparison star we have used an independent quadratic polynomial model to account for systematic trends in the light curves. Fig.~\ref{fig:starcomp} presents the data obtained from the differential photometry of HATS-24 with respect to each comparison star used in our analysis. Fig.~\ref{fig:starcomp} also presents the fit models (green line), which include the trend models and the global transit model. Notice that each light curve has the transit signature and, in addition, a trend that is well modeled by a quadratic polynomial of the form  $P_j (t) = a_{j}t^{2} + b_{j}t + c_{j}$, for $j$ being the index for each comparison star. All the coefficients $a_{j}$, $b_{j}$, and $c_{j}$ are obtained by Bayesian inference analysis simultaneous to the transit parameters.

\begin{figure}
\centering
\includegraphics[width=0.45\textwidth]{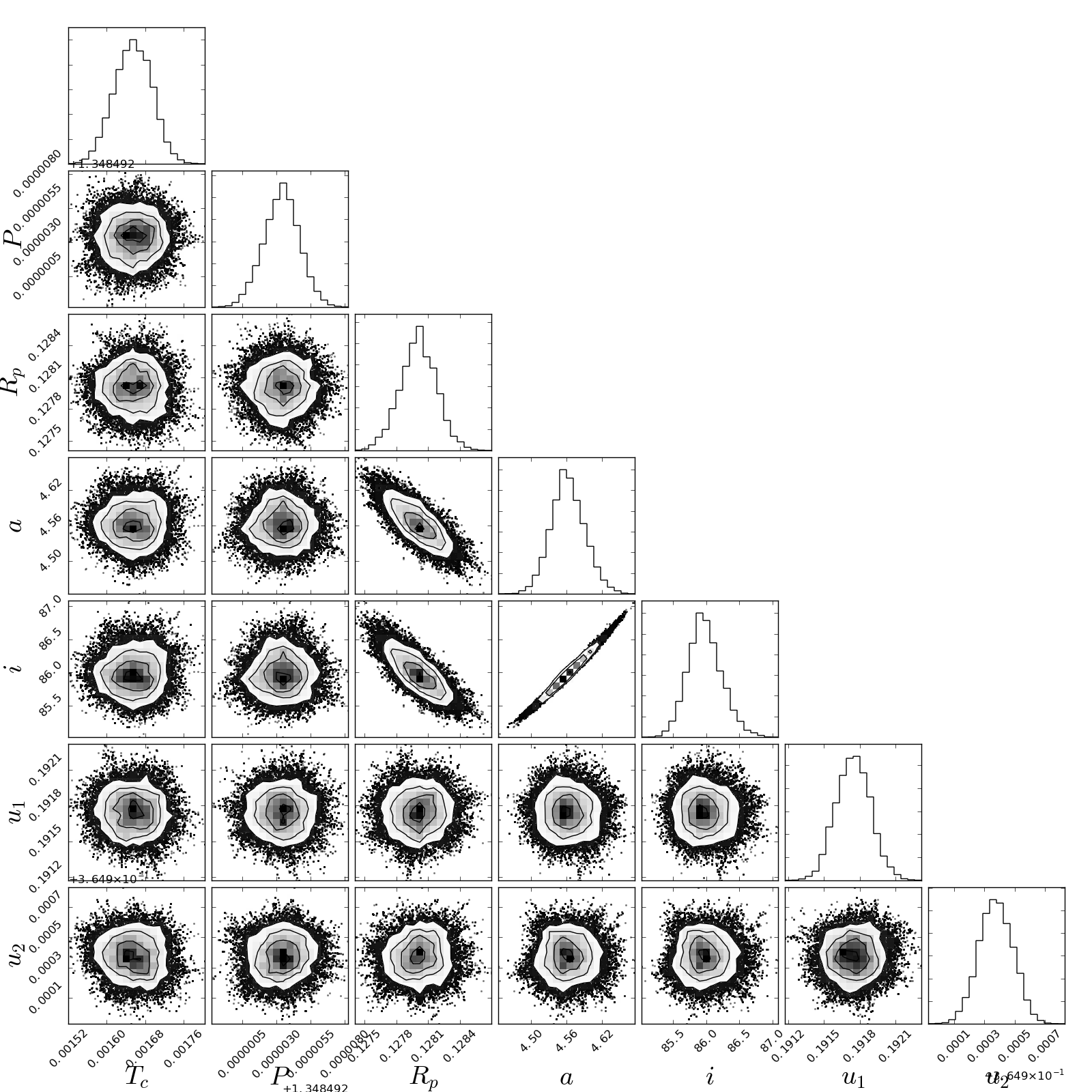}
\caption{MCMC diagnostics: pairwise plots of the parameters in the posterior distribution obtained from the last 1000 iterations of the MCMC sampler; the panels in the main diagonal show the one-dimensional projection of the posterior probability distribution for each parameter of the transit model.}
\label{fig:parsplot}
\end{figure}

To determine the posterior probability distribution of each parameter, 5000 iterations were performed through the MCMC sampler, where the initial 4000 iterations have been discarded as burn-in. Fig.~\ref{fig:parsplot} presents the MCMC diagnostic plots, i.e., the pairwise plots of the parameters in the posterior distribution and the one-dimensional projection of the posterior probability distribution for each transit model parameter. This analysis was also obtained for the parameters of the trend polynomials of each comparison star, but for simplicity we only show here the parameters of the transit.

The final value of each parameter is calculated as the 50th percentile of the posterior distribution generated by the MCMC, and the uncertainties are calculated as the mean value between the 50th percentiles minus 16 and the 84th percentiles minus 50. Table~\ref{table:finalfit} presents the final results for the transit parameters.

{\renewcommand{\arraystretch}{1.35}
\renewcommand{\tabcolsep}{0.1cm}
\begin{table}
\centering
\centering \caption{Final fit parameters of HATS-24b.}
\label{table:finalfit}
\begin{tabular}{cccc}
\hline
Parameter       & Value                        &  Unit   \\
\hline
$T_{c}$      & 2457948.709321   $_{-0.000038}^{+0.000041}$    & BJD \\
$P$       &  1.34849540    $_{-0.0000012}^{+0.0000013}$   & days\\
$R_{p}/R_{\star}$  &  0.12801   $\pm$  0.00017    & -\\
$a_{p}/R_{\star}$   &  4.561      $_{-0.030}^{+0.027}$      & -\\
$i$      &  85.97        $_{-0.28}^{+0.23}$      & degrees\\
$e$      &  0                              &  - \\
$\omega$   &  90                             & degrees\\
$u_{1}$            &  0.19187           $\pm$  0.00017   & - \\
$u_{2}$            &  0.36543           $\pm$  0.00009   & - \\
\hline
\end{tabular}
\end{table}
}

\begin{figure}\centering
\includegraphics[width=0.45\textwidth]{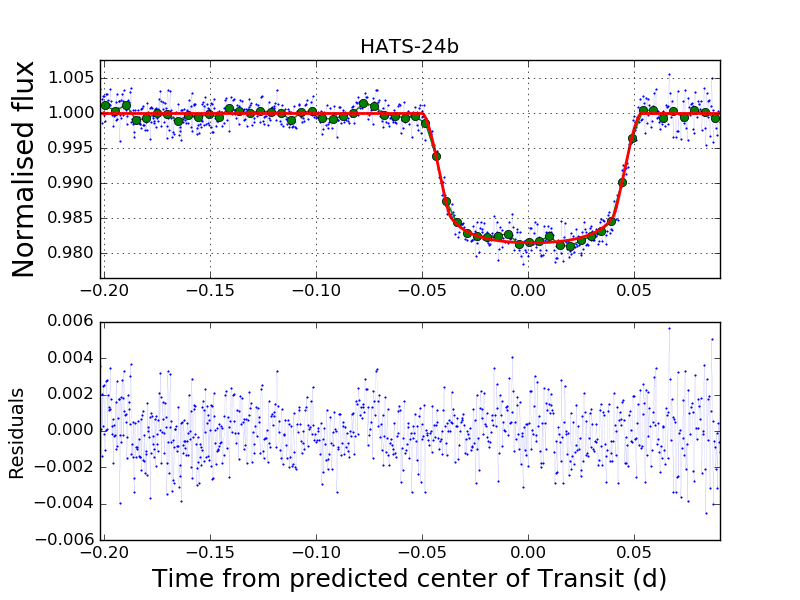}
\caption{Top panel shows the reduced light curve of HATS-24 (blue dots), the binned data (green circles), and the fit transit model (red line). Bottom panel shows the residuals.}
\label{fig:fit}
\end{figure}

\begin{figure}\centering
\includegraphics[width=0.45\textwidth]{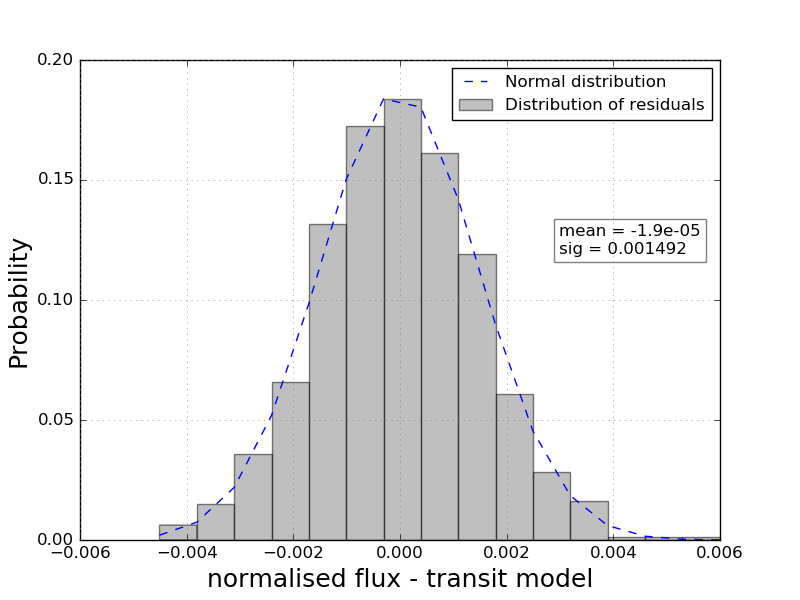}
\caption{Probability distribution of residuals (gray bars), and the fit normal distribution (blue dashed line), N($\mu$, $\sigma$), where $\mu = 2\times10^{-6}$ and $\sigma=0.001490$.}
\label{fig:hist}
\end{figure}

Fig.~\ref{fig:fit} shows the final results of our analysis, where we present the mean of all trend-subtracted light curves of HATS-24 (blue dots), the transit model (red line) using the fit parameters from Table~\ref{table:finalfit}, and the binned data (15 points per bin) with uncertainties calculated by the standard deviation.  Fig.~\ref{fig:hist} shows the probability distribution of residuals, where we fit a normal distribution N($\mu$, $\sigma$). We obtained $\mu = 2\times10^{-6}$ and $\sigma=0.00149$. Notice that the distribution of residuals is well described by a normal distribution, with zero mean and a standard deviation of $\sigma=0.15\%$.

\vspace{-5pt}

\subsection{Ephemeris}

Using the results presented in Section \ref{sec:results} for the central time of transit ($T_{c}$), and combining the information available in the literature for two other transit events of HATS-24b, we have obtained an improved ephemeris for the transits of this exoplanet. We have found two reported transits of HATS-24b in the literature; one from \citet{Bento2017}, and another by Phil Evans published in the catalog Exoplanet Transit Database \citep{poddany2010}. Table~\ref{table:efemeride} shows the details of these three observed transits of HATS-24b.

\begin{table*}
\centering
\centering \caption{Observed transits of HATS-24~b}
\label{table:efemeride}
\begin{tabular}{ccccc}
\hline
T$_{c}$                        & Epoch &    O - C                       &  Duration of Transit\footnotesize$^a$ & Reference   \\
(BJD $-$ 2457000)              &       &   (days)                       &      (minutes)                     &              \\
\hline
038.47327   ${^+_-}$ 0.00038   & -675 & -0.000048 ${^+_-}$ 0.00038   & 145.2 ${^+_-}$ 2.2   & \citet{Bento2017}   \\
608.89160   ${^+_-}$ 0.00096   & -252 &  0.0037   ${^+_-}$ 0.00096     & 142.2 ${^+_-}$ 3.3   & Phil Evans\footnotesize$^b$ \\
948.7093208 ${^+_-}$ 0.000039 &  0   & -0.0000079${^+_-}$ 0.000039 & 146.3 ${^+_-}$ 1.4    & This Work       \\
\hline
\multicolumn{5}{l}{\footnotesize$^a$ The duration of transit is calculated using the relation presented in \cite{winn2010}}\\
\multicolumn{5}{l}{\footnotesize$^b$ \url{http://var2.astro.cz/tresca/transit-detail.php?id=1470685388}}\\
\end{tabular}
\end{table*}

\begin{figure}
\includegraphics[width=1.05\columnwidth]{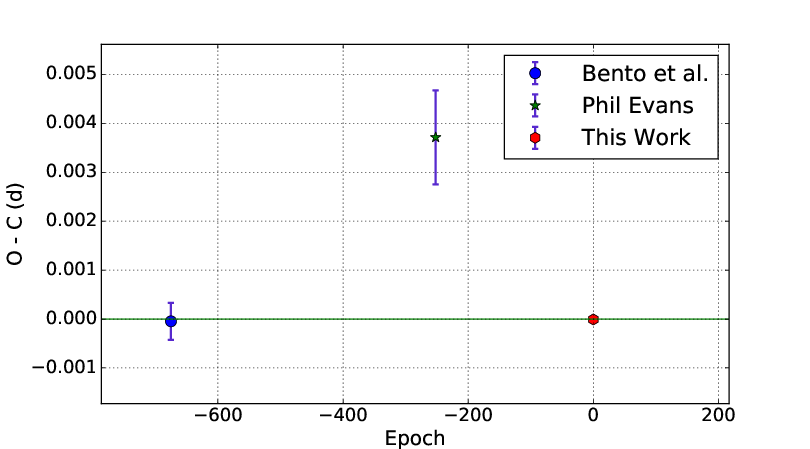}
\caption{Observed minus computed (O-C) diagram of the central times of transits as a function of epoch for the three transit events of HATS-24b.}
\label{fig_o-c}
\end{figure}

We have used \texttt{GaussFit} \citep{Jefferys1988} to perform a linear fit to the data presented in Table~\ref{table:efemeride}. The model is given by $T_{c} = T_{0} + E\times P$, for $E$ being the epoch from the time of the transit reported in this work, $T_{0}$ being the central time of transit at initial epoch $E=0$, and $P$ being the orbital period. \texttt{GaussFit} performs a least squares robust analysis which takes into account the errors in the observations. We obtained the following ephemeris for the central time of transit:  

\begin{equation}
T_{c} = (2457948.709321 \pm 0.000039) + E (1.3484978 \pm 0.0000009),
\end{equation}

where we have adopted the value of $T_{0}$ as our measured value of $T_{c}$ because it is more precise than the fit value of $T_{0}$. The orbital period obtained in this analysis represents the most precise value at the moment. Fig.~\ref{fig_o-c} presents the observed minus computed (O-C) values for the central time of transit ($T_{c}$) as a function of epoch. Notice that the value of $T_{c}$ measured by Phil Evans is more than $3\sigma$ away from the predicted value. We believe that this is likely due to a systematic error not accounted for in his measurements, otherwise it should represent some variation in the central time of transit. To confirm the latter hypothesis it requires further follow-up of the transits of HATS-24b. A possible variation in the central time of transit may indicate the presence of additional planets in the system.  

\subsection{Stellar Parameters}
\label{sec:spec}

We have used the high resolution spectrum of HATS-24, which was kindly provided by the authors of \cite{Bento2017}, in addition to the Vesta spectrum obtained from the ESO archive (program id 090.D-0133(A)) to recalculate the stellar parameters of HATS-24. Both spectra were observed with the FEROS spectrograph \citep{Kaufer1999}, where the spectrum of HATS-24 has a maximum signal-to-noise ratio (SNR) of $\sim$40, and the Vesta spectrum has SNR$\sim$115. 

The stellar parameters were initially obtained through the standard spectroscopic methods, i.e., through the excitation and ionization equilibrium using the equivalent widths (EW) of Fe I and Fe II lines \citep{mel09, ram14, tuc14, tuc16}. This is done to ensure that the atmospheric model returns the same iron abundances for every line, independently of its excitation potential, ionization state or line strength.
Since HATS-24 is a solar type star, we can use the solar spectrum, obtained from the reflected light of Vesta, as reference for the differential analysis. The EWs were measured manually with the task \texttt{splot} in \texttt{IRAF}, by fitting Gaussian profiles, and the abundances were determined using the 2014 version of the LTE code \texttt{MOOG} \citep{sne73} adopting the MARCS grid of 1D-LTE model atmospheres \citep{gus08}. Since the spectra available have poor SNR for this kind of analysis, we decided to use the trigonometric gravity, $\log{g}=4.37 \pm 0.045$~cm~s$ˆ{-2}$, obtained with the apparent magnitude from Simbad \citep{wenger2000} and the parallax from GAIA \citep{gaia2018}. By fixing the gravity value, we parse through stellar models to find the best fit values of effective temperature (T$_{\rm eff}$), metallicity ([Fe/H]), and microturbulence velocity ($\nu$).

We employed the \texttt{Python} code \texttt{q2}\footnote{\url{https://github.com/astroChasqui/q2}} \citep{ram14} to determine the chemical abundances, by calling \texttt{MOOG} routines, to perform the necessary differential analysis. The uncertainties in the stellar parameters were determined taking into account the observational (due to uncertainties in the EW measurement) and systematic (from the degeneracy of the stellar parameters) uncertainties.

\begin{figure}
\centering
\includegraphics[width=0.8\columnwidth]{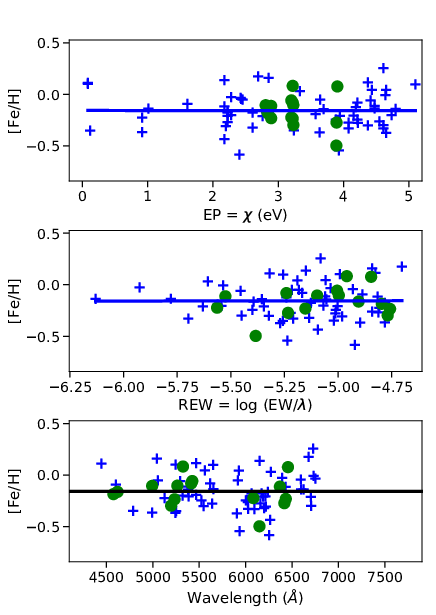}
\caption{Excitation and ionization equilibrium of Fe abundances for HATS-24 using the solar spectrum as reference star. Crosses represent Fe I and filled circles Fe II.}
\label{exc1}
\end{figure}

Fig.~\ref{exc1} shows the excitation and ionization equilibrium of Fe I and Fe II lines. The final solution for the stellar surface parameters of HATS-24 determined by the EW differential analysis are presented in Table \ref{stel_param}.

\begin{table}
\centering
\caption{Stellar surface parameters obtained for HATS-24 determined with EW diferential analysis.}
\label{stel_param}
{\centering
\begin{tabular}{cccc}
\hline
  Parameter & Symbol & HATS-24 & Unit \\
 \hline
 \hline
effective temperature & T$_{\rm eff}$    & 6125 $\pm$ 94     & K \\
surface gravity & $\log{g}$      & 4.370 $\pm$ 0.045 & cm~s$ˆ{-2}$ \\
metallicity & $[$Fe/H$]$ & -0.229 $\pm$ 0.058  & dex \\
microturbulence velocity & $\nu$           & 1.75 $\pm$ 0.15   & km\,s$^{-1}$\\
 \hline
 \end{tabular}
 }
\end{table}

\begin{figure}
\centering
\includegraphics[width=1.0\columnwidth]{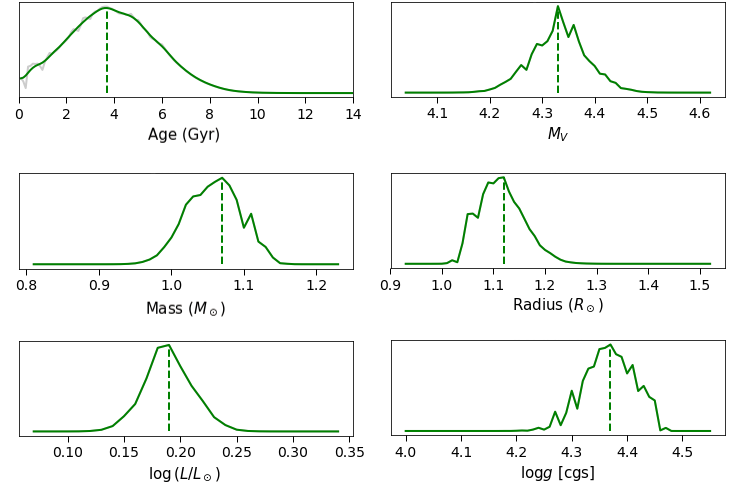}
\caption{Green solid lines show the probability distribution for HATS-24 parameters based on Yonsei-Yale isochrones. Green dashed lines show the mode of each distribution.}
\label{host_param_prob}
\end{figure}

\begin{figure}
\centering
\includegraphics[width=0.9\columnwidth]{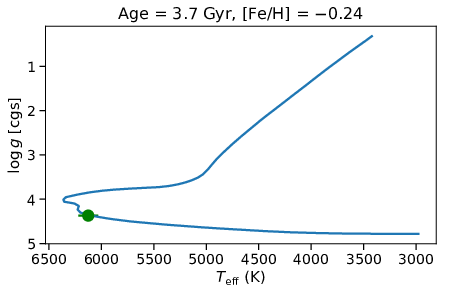}
\caption{Blue line shows the YY isochrone in the $T_{\rm eff}$ vs. $\log{g}$ space for an age of 3.7~Gy and metallicity of $[$Fe/H$]$=-0.24~dex. Green point shows the location of HATS-24, where we have used the fit parameters from Table \ref{stel_param}. }
\label{hats24_isonea}
\end{figure}

 The age, luminosity, absolute magnitude, mass, and radius, presented on Table~\ref{host_param}, have been calculated using the modified Yonsei-Yale (YY) isochrones \citep{yi01}, as described in \cite{ram13, ram14}. These parameters are estimated from the probability distribution functions shown in Fig.~\ref{host_param_prob}. For further calculations using the stellar parameters we have adopted the mode of the distribution of each parameter (green dashed lines in Fig.~\ref{host_param_prob}). Fig.~\ref{hats24_isonea} presents the most likely isochrone (age of 3.7~Gy and metallicity of $[$Fe/H$]$=-0.24~dex) that fits the parameters of HATS-24 obtained from our analysis.

\begin{table}
\centering
\centering \caption{HATS-24 parameters estimated with the isochrones method.}
\label{host_param}
\begin{tabular}{cccccc}
\hline
Parameter   & Symbol     &  Mode  & Mean & Unit \\
\hline 
Age      & - & 3.7$_{-1.8}^{+2.0}$  & 3.9 $\pm$1.8   & Gy\\
Luminosity & $\log{L_{\star}}$   & 0.190$\pm$0.021   & 0.190$\pm$0.021   & $L_{\odot}$ \\
Abs. mag & $M_{V}$    & 4.330$\pm$0.051   & 4.332$\pm$0.053 & mag \\
Mass & $M_{\star}$      & 1.070$_{-0.046}^{+0.032}$   & 1.058$\pm$0.037  & $M_{\odot}$\\
Radius & $R_{\star}$     & 1.120$_{-0.047}^{+0.044}$   & 1.115$\pm$0.043     & $R_{\odot}$\\
\hline
\end{tabular}
\end{table}

\subsection{Derived Parameters of HATS-24b}
\label{devparams}

Combining the transit parameters presented in Section \ref{sec:transitpars} with the stellar parameters presented in Section \ref{sec:spec} we were able to calculate the planetary parameters presented in Table~\ref{table:derivedparams}. The equilibrium temperature was calculated as in \cite{Heng2013}, where we assumed an albedo of $A=0.1$, and an uniform heat redistribution, i.e., $f=1/2$. 

\begin{figure}
\centering
\includegraphics[width=1.1\columnwidth]{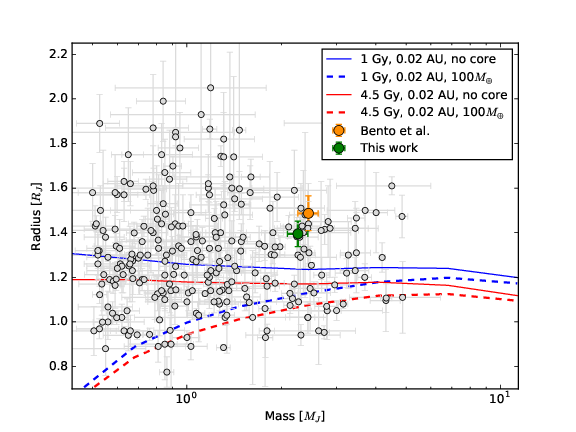}
\caption{Mass-radius relation for hot Jupiters, where we include exoplanets with masses in the range $0.5\,M_{\rm J}<M_{\rm p}<5\,M_{\rm J}$ and with periods shorter than $10$ days (source: NASA Exoplanet Archive catalog). The orange point represents the previous result from \citet{Bento2017} for HATS-24b and the green point represents our result. We present theoretical models for planet structures from \citet{Fortney2007} where in blue are the 1Gy models and in red are the 4.5Gy models. Both models are presented for both no core (solid lines) and 100$M_{\oplus}$ core (dashed lines) scenarios.}
\label{Mass-radius}
\end{figure}

Figure \ref{Mass-radius} presents the mass-radius relation for hot Jupiters, where we included exoplanets with masses in the range $0.5\,M_{\rm J}<M_{\rm p}<5\,M_{\rm J}$ and with periods shorter than $10$ days. We present theoretical models for planet structures obtained from \cite{Fortney2007}. Notice that our measurements confirm that HATS-24b is an inflated hot Jupiter, however the new estimation for the radius is now less than 3$\sigma$ above the model for a pure helium-hydrogen planet at approximately 1~Gy.

\begin{table}
\centering
\centering \caption{Derived parameters for HATS-24b}
\label{table:derivedparams}
\begin{tabular}{cccc}
\hline
Parameter        & Value  & Unit \\
\hline
$M_{p}$      & 2.26  $\pm$ 0.17   & M$_{\rm J}$\\
$R_{p}$      & 1.395 $\pm$ 0.057  & R$_{\rm J}$\\
$\rho_{p}$   & 1.03  $\pm$ 0.15   & g~cm$^{-3}$ \\
$a_{p}$      & 0.0238$\pm$ 0.0010 & AU\\
$T_{\rm eq}$     & 2166  $\pm$ 53     & K\\
\hline
\end{tabular}
\end{table}

\section{Conclusions}
\label{sec:conclusions}

We have presented observations of a transit event of the exoplanet HATS-24b, where we have used for the first time one of the new CCD cameras of the SPARC4 instrument. We present differential photometry time-series covering an entire transit event of HATS-24b. We obtained a final photometric precision of $0.15\%$ for a V=12.83~mag star.  This confirms that SPARC4 will be an excellent instrument for relative high-precision photometry with the potential for the detection and characterization of transits of hot Jupiters. Notice that we obtained the transit depth of HATS-24b with precision of 170~ppm. The four simultaneous channels of SPARC4 will permit observations of transits simultaneously in four photometric bands, allowing one to constrain models of transmission spectroscopy of transiting planets.

We have presented a robust approach using Bayesian statistical inference to determine the planetary parameters with better precision compared to previous results from \cite{Bento2017}. The improved precision for the central time of transit combined with data from the literature allowed us to calculate a new ephemeris for the transits of HAT-24b, improving the orbital period and the prediction of future transit events for this object.  Finally, we have used available high resolution spectra from HATS-24 and Vesta observed with the FEROS spectrograph to perform a differential analysis to obtain the stellar parameters of HATS-24. We have found an effective temperature of $6125\pm94$~K, confirming that this star is likely a G0 type with an age of $3.9\pm1.8$~Gyr, relatively older than previous measurements.  Combining the stellar analysis with our fit transit parameters we obtained a new estimate for the following parameters of the planet: mass, radius, semi-major axis, equilibrium temperature, and density. Our results are more consistent with the theoretical models for planet structures, although the radius of HATS-24b still appears to be inflated compared to any of the models presented.

\section*{Acknowledgements}

We would like to thank the LNA/OPD staff for their support during the observations, and for promptly responding to our requests. J. M. Oliveira acknowledges the financial support from CNPq agency under the PCI project, process number 454781/2015-6. We would also like to thank J. Bento for providing the spectroscopic data of HATS-24.



\end{document}